# ZnO and ZnO$_{1-x}$ based thin film memristors: The effects of oxygen deficiency and thickness in resistive switching behavior

Fatih Gul [a, 1, *]

[a] Department of Electrical & Electronics Engineering, Ataturk University, Erzurum 25240, Turkey

Hasan Efeoglu [a, b]

[b] Nano-science and Nanoengineering Research and Application Center, Ataturk University, Erzurum 25240, Turkey

**Abstract**

In this study, direct-current reactive sputtered ZnO and ZnO$_{1-x}$ based thin film (30 nm and 300 nm in thickness) memristor devices were produced and the effects of oxygen vacancies and thickness on the memristive characteristics were investigated. The oxygen deficiency of the ZnO$_{1-x}$ structure was confirmed by SIMS analyses. The memristive characteristics of both the ZnO and ZnO$_{1-x}$ devices were determined by time dependent current–voltage (I-V-t) measurements. The distinctive pinched hysteresis I-V loops of memristors were observed in all the fabricated devices. The typical homogeneous interface and filamentary types of memristive behaviors were compared. In addition, conduction mechanisms, on/off ratios and the compliance current were analyzed. The 30 nm ZnO based devices with native oxygen vacancies showed the best on/off ratio. All of the devices exhibited dominant Schottky emissions and weaker Poole-Frenkel conduction mechanisms. Results suggested that the oxygen deficiency was responsible for the Schottky emission mechanism. Moreover, the compliance currents of the devices were related to the decreasing power consumption as the oxygen vacancies increased.

## 1. Introduction

A memristor (memory + resistor) is a fourth fundamental circuit device which functions as a passive two-terminal non-linear device as well as a resistor (R), capacitor (C) and inductor (L). The device was first suggested by Leon Chua, in 1971 [1], and a TiO$_2$-based physical example and a linear model of a memristor was presented in 2008 [2]. Thenceforth, the memristor has attracted increasing attention. There have been several application areas suggested for memristors or memristive systems including

[1] Corresponding author,
E-mail address: fatihgul@atauni.edu.tr (F. Gul).
[*] Permanent address: Department of Electrical & Electronics Engineering, Gumushane University, Gumushane 29100, Turkey

memory devices [3], neuromorphic networks [4], logic circuits [5] and others [6]. Due to the potential of an ultra-high density memory device with lower energy consumption and faster switching speed, researchers have focused great attention on non-volatile memory applications for the memristor. Therefore, memristors seem quite viable for technology in the near future [7].

The pinched hysteresis loops serve as a fingerprint in the characterization of memristors [8]. As in resistive switching devices, a typical pinched hysteresis loop is seen at the 1st and 3rd quadrants of the current-voltage (I-V) curves [9]. Put differently, all memristor devices show resistive switching (RS) characteristics, regardless of the material or operating mechanisms [10].

In the explanation of time dependent current–voltage (I-V-t), there are two types of physical switching mechanisms based on molecular or ionic models: the homogeneous interface-type and the filamentary (conduction path) type [9,11,12]. In the homogeneous type, resistance change is the result of the migration of oxygen vacancies as the major carriers [13,14]. The filament-type mechanism is associated with the formation and rupture of conductive filaments in the metal oxide layers [9,14].

Most of the metal oxides have homogeneous and/or filamentary characteristics and exhibit RS memristive behavior [9,15,16]. In recent decades, amongst these oxide-based materials, ZnO has become increasingly popular for use in memristor-based devices [17,18] owing to its distinctive properties [19,20].

In this study, the homogeneous and the filamentary types of memristive behavior of Al/ZnO/Al and Al/ZnO/ZnO$_{1-x}$/ZnO/Al -based memristors were investigated. The two types of switching behavior of the devices were expounded by I-V-t characterization. In addition, the conduction mechanisms, on/off ratios and compliance currents of the devices were compared.

**2. Experimental details**

Thin films of ZnO and ZnO$_{1-x}$ (30 nm and 300 nm in thickness) were grown by DC reactive magnetron sputtering using a metallic Zn target (Kurt Lesker 2" Ø, 99.995% purity) in argon (Ar) and oxygen (O$_2$) gas flow. The base pressure of the chamber was lower than $2 \times 10^{-6}$ Torr and the Ar to O$_2$ ratio was 20:1. The substrates were not heated and a voltage of -350 VDC was used during the sputtering process. Film thickness was monitored by the INFICON XTM/2. The other parameters used for the samples are given in **Table 1**. All ZnO thin films were grown under the same conditions, with the aim of ensuring comparable and reproducible samples.

**Table 1**. Growth parameters of the ZnO and ZnO$_{1-x}$ devices

| Device | Thickness | Material | Ar | $O_2$ |
|---|---|---|---|---|
| ZnO | 30 nm | ZnO | 20 sccm | 1 sccm |
| | 15 nm | ZnO | 20 sccm | 1 sccm |
| ZnO/ZnO$_{1-x}$/ZnO | 1 nm (Estimated) | ZnO$_{1-x}$ | 20 sccm | Cut-off for 10 Seconds |
| | 15 nm | ZnO | 20 sccm | 1 sccm |
| ZnO | 300 nm | ZnO | 20 sccm | 1 sccm |
| | 150 nm | ZnO | 20 sccm | 1 sccm |
| ZnO/ZnO$_{1-x}$/ZnO | 10 nm (Estimated) | ZnO$_{1-x}$ | 20 sccm | Cut-off for 60 Seconds |
| | 150 nm | ZnO | 20 sccm | 1 sccm |

Standard RCA cleaned p$^{++}$ Si wafers were oxidized at 1050 °C under a dry $O_2$ flow rate of 10 sccm for 30 min to obtain the $SiO_2$/p$^{++}$Si used as the substrate for device fabrication and characterization. The top and bottom Al electrodes were completed using physical vapor deposition (PVD) under vacuum conditions of lower than 1 mTorr. The top Al electrode metallization was done using a shadow mask with 1 mm holes (**Figure 1**).

Memristive (I-V-t) characterizations were performed by a Keithley 2400 SourceMeter® and custom-designed memristor characterization software at room temperature using a probe station. Secondary ion mass spectrometry (SIMS) analyses of the samples were achieved using the Hiden® SIMS workstation.

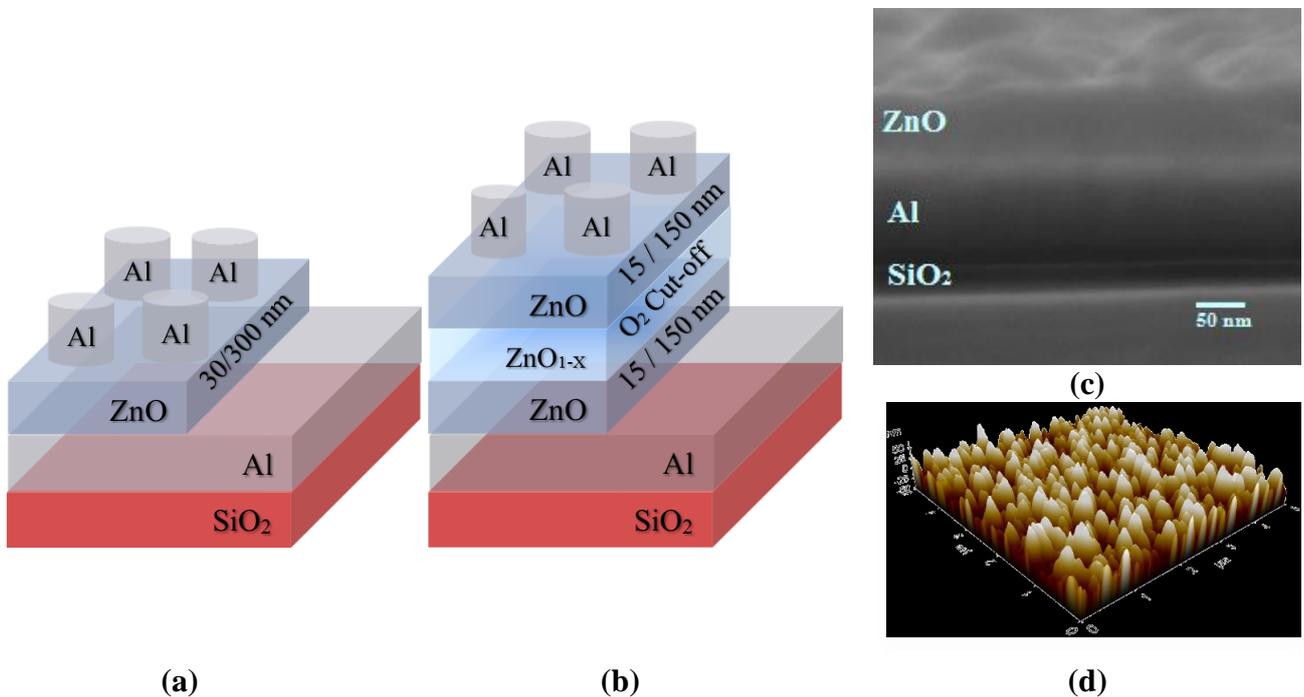

**(a)**          **(b)**          **(d)**

**Figure 1.** Schematic representation of **(a)** the Al/ZnO/Al device and **(b)** the Al/ZnO/ZnO$_{1-x}$/ZnO/Al device. Cross-section image of the ZnO structures **(c)**. Surface morphology AFM image of the ZnO **(d)**.

## 3. Results and Discussion

The optical and the structural analyses of the ZnO were confirmed by ultraviolet-visible (UV-Vis) spectrophotometry and X-ray powder diffraction (XRD), respectively [21]. **Figure 1(d)** shows AFM image of the ZnO surface of the device. The scan rate was 0.44 Hz, the scan configuration was 512 × 512 pixels, and the scan size was 5 × 5 μm². It is confirmed that the surface was formed uniformly having the surface roughness factor $R_a$=0.044 μm. The cross-section SEM image of the Al/ZnO/SiO$_2$ structure showed in the **Figure 1(c)**.

The effect of cutting the oxygen flow midway in the growing process of the ZnO was examined via SIMS. The secondary ion count as a function of depth of the ZnO and ZnO$_{1-x}$ structures is depicted in **Figure 2.** It was obvious that the intentional cutting of the O$_2$ gas flow decreased the O$_2$: Zn and ZnO: Zn ratios through the depth of 30 nm.

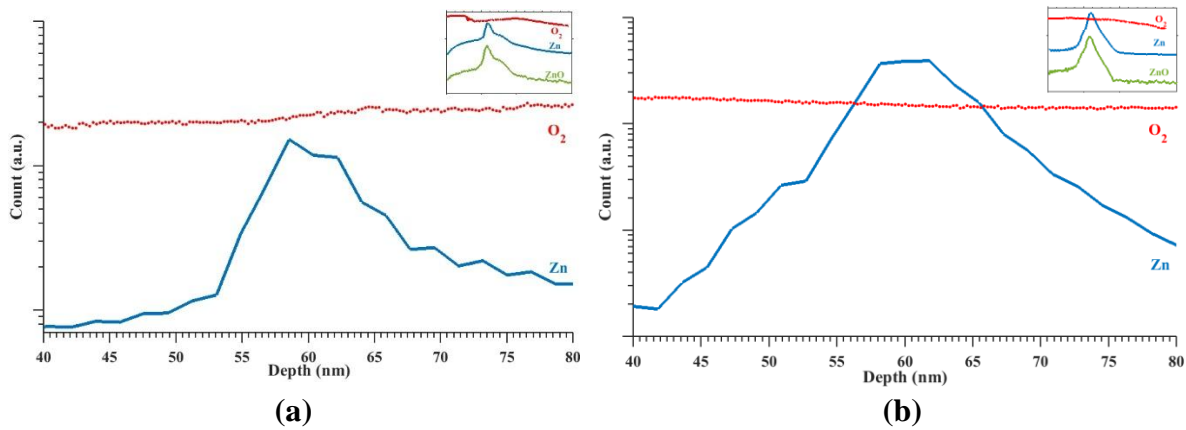

**Figure 2.** SIMS profiles of **(a)** ZnO and **(b)** ZnO$_{1-x}$ thin films on SiO$_2$ substrates (O$_2$, Zn and ZnO depth profiles are shown in inset graphs)

The most common method of defining resistive switching behavior is I-V-t measurement with a pre-defined voltage scan rate. The direct current I-V-t characterizations of the ZnO and ZnO$_{1-x}$ based resistive switching memristor devices were done by means of specially designed software for I-V-t measurement using a Keithley 4200 SourceMeter®. In order to avoid permanent damage from excess current, the compliance current (CC) must be set. The compliance current also defines the working power consumption current of a memristor device. Therefore, the CC was determined for each device, as demonstrated in **Table 2**. The resistive switching behavior of the Al/ZnO/Al and Al/ZnO/ZnO$_{1-x}$/ZnO/Al devices was detected by changing the voltage between -2.5 V and +2.5 V with a sweep rate of 0.1 V/s.

**Table 2**. Comparisons of resistive switching behavior

| Device | | Al/ZnO/Al | Al/ZnO/ZnO$_{1-x}$/ZnO/Al | Al/ZnO/Al * | Al/ZnO/ZnO$_{1-x}$/ZnO/Al |
|---|---|---|---|---|---|
| Thickness | | 30 nm | 30 nm | 300 nm | 300 nm |
| Require Forming | | √ | N/A | √ | N/A |
| CC | | 2 mA | 0.5 mA | 3 mA | 0.5 mA |
| Switching Mechanism | | Filamentary | Homogeneous | Homogeneous | Filamentary |
| On/Off Ratio (@0.1V) | | 54.8 | 3.48 | 2.05 | 4.8 |
| Median Resistance | HRS (KΩ) | 16 | 17.6 | 3.2 | 41.2 |
| | LRS (KΩ) | 0.3 | 5.06 | 1.6 | 8.6 |

| Conduction Mechanism | Low Electric Field Region | Ohmic | Ohmic | Ohmic | Ohmic |
| --- | --- | --- | --- | --- | --- |
| | High Electric Field Region | SE >P-F>SCLC | SE >P-F>SCLC | SE >P-F>SCLC | SE >P-F>SCLC |
| Changing State with loops | | HRS to LRS | LRS to HRS | HRS to LRS | LRS to HRS |
| Fitting Slope for SE | | 4.7 | 6.6 | 5.1 | 5.5 |
| Fitting Slope for P-F | | 2.7 | 4.5 | 2.9 | 3.2 |

* The column values were recalculated from a previous work for comparison [21].

The pinched hysteresis I-V loop as shown in **Figure 3** is the first characteristic fingerprint of the memristor [8]. The memristor has two states when used as a resistive switch: the ON (SET) state or LRS (low-resistance state) and the OFF (RESET) state or HRS (high-resistance state) [12].

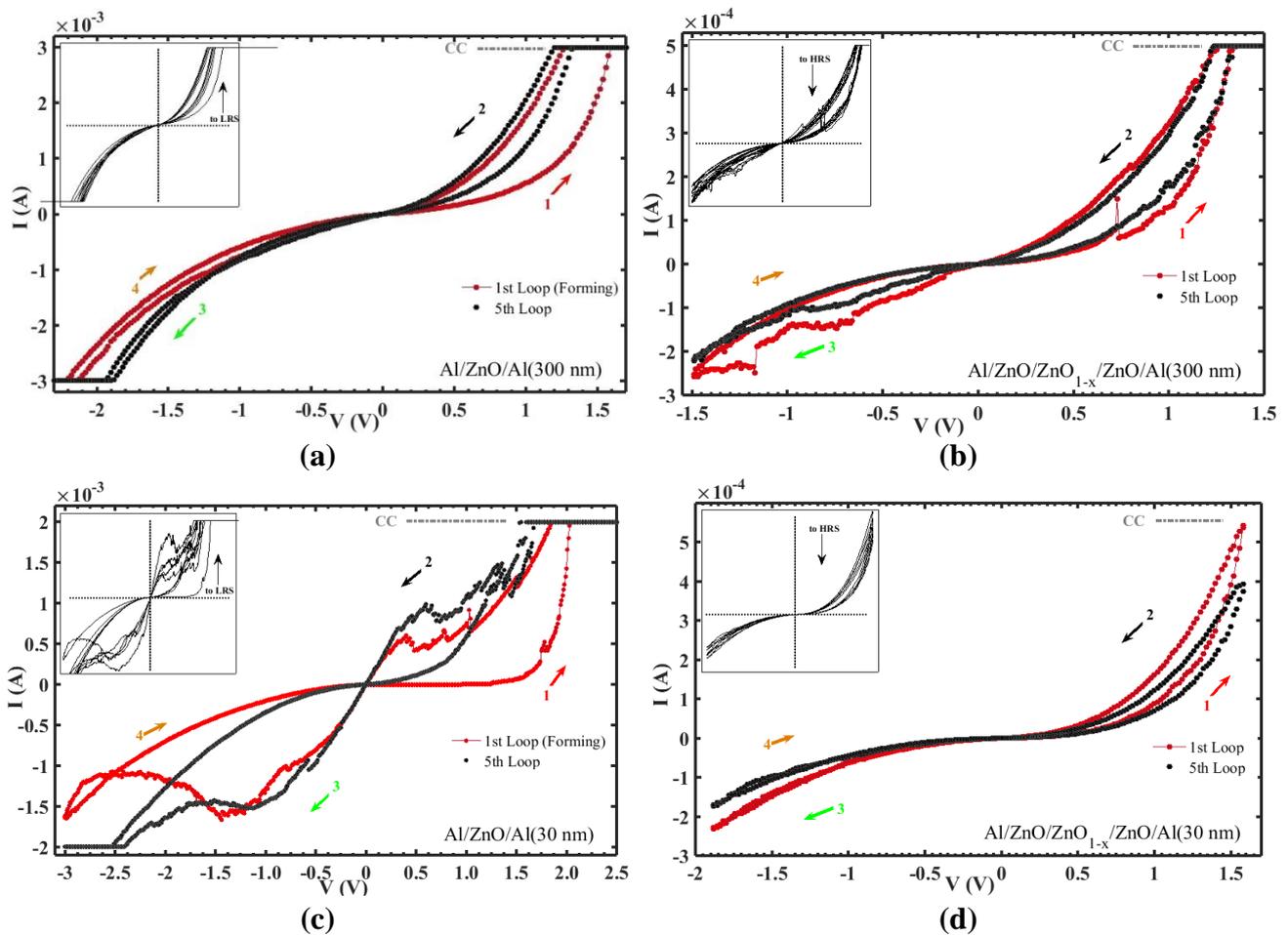

**Figure 3.** Hysteretic loops of memristor devices for the 1st and 5th loops: **(a)** Al/ZnO/Al (300 nm thick), **(b)** Al/ZnO/ZnO$_{1-x}$/ZnO/Al (300 nm thick), **(c)** Al/ZnO/Al (30 nm thick), **(d)** Al/ZnO/AlZnO$_{1-x}$/ZnO/Al (30 nm thick) (Inset graphs show other continuously repeated successful loops and shifting HRS/LRS states of the memristor devices)

The bipolar and unipolar switching actions of memristors can be categorized according to current I-V characteristics. In bipolar switching, characteristics are resolved by amplitude and polarity of the applied voltage, whereas unipolar switching depends only on the amplitude of the applied voltage [9]. All of the presented devices showed symmetric bipolar RS characteristics, regardless of the thickness and the level of the oxygen vacancies.

In the explanation of the I-V-t there are two types of physical switching mechanisms, based on molecular or ionic models: the homogeneous interface-type and the filamentary (conduction path) type [9,11,12]. In the homogeneous type, the migration of oxygen vacancies as the majority carriers causes change of resistance [13,14]. The filament-type mechanism is associated with the formation and rupture of conductive filaments in the metal oxide layer [9,14]. Both types of mechanisms were observed in the devices presented in this study. The filamentary-type switching was observed in the 300 nm-thick Al/ZnO/ZnO$_{1-x}$/ZnO/Al structure and in the 30 nm-thick Al/ZnO/Al device (**Figure 3 (b),(c)**). On the other hand, the homogeneous-type switching was observed in the 300-nm thick Al/ZnO/Al device and the 30 nm-thick Al/ZnO/ZnO$_{1-x}$/ZnO/Al structure (**Figure 3 (a),(d)**). This suggested that the homogeneous switching mechanism increased as the oxygen vacancies increased in the thinner devices, whereas the filamentary mechanism increased as the thickness and oxygen vacancies increased. It is suggested that the oxygen vacancies cause the conductive filament route in the ZnO structures [22].

Memristor devices sometimes need a forming process to trigger the RS characteristics [12]. The presented 300 nm-thick Al/ZnO/Al device and the 30 nm-thick Al/ZnO/Al device both needed forming processes (**Figure 3 (a),(c)**). However, the devices with oxygen-deficient Al/ZnO/ZnO$_{1-x}$/ZnO/Al in both 30 nm and 300 nm thicknesses (**Figure 3 (b),(d)**) did not need forming cycles. This suggested that the forming processes were related to the oxygen deficiency level in the memristor structures, regardless of the thickness. In other words, increased oxygen deficiency diminished the initial forming process requirement.

The other issue of memristive behavior is the shifting of states from HRS to LRS or from LRS to HRS after application of the switching cycles [23]. In the ZnO$_{1-x}$ based structures which intentionally contained oxygen vacancies, both the 30 nm- and the 300 nm-thick devices shifted states from LRS to HRS after the switching cycles were applied (inset of **Figure 3 (b), (d)**). However, the devices containing only native oxygen vacancies shifted from HRS to LRS after the continuous cycles were applied (inset of **Figure 3 (a),(c)**). The increase of the HRS resistance with the cycling can be explained by the degrading of the oxygen vacancies as carriers [23].

The I-V curves of the Al/ZnO/Al and Al/ZnO/ZnO$_{1-x}$/ZnO/Al devices were plotted on a double logarithmic scale with the aim of recognizing the carrier transport mechanisms (**Figure 4 (a),(d)** and **Figure 5 (a),(d)**). The low electric field region in all devices for both the LRS and HRS states can be fitted linearly, and all slopes are close to 1, which demonstrates that the ohmic conduction mechanism is compatible with findings in previous ZnO-based memristor studies [15,24,25].

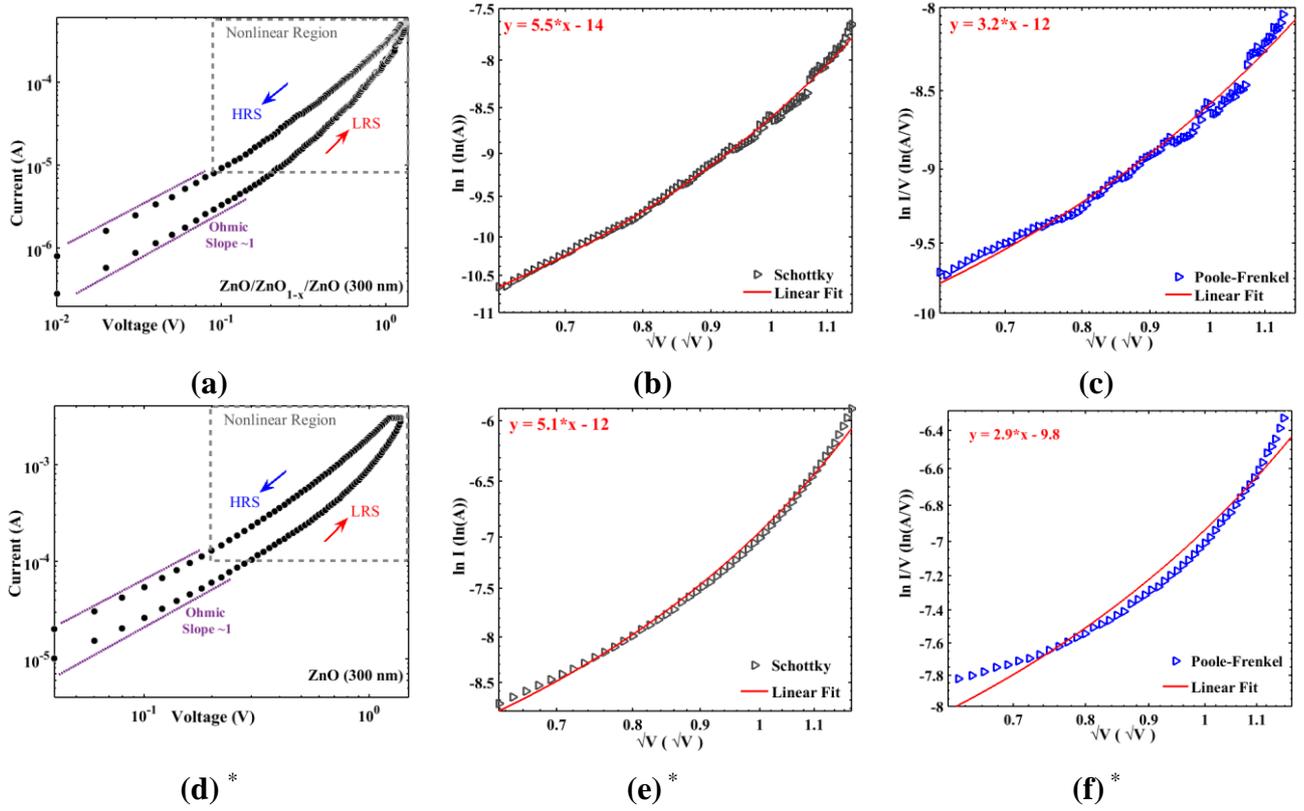

*Re-plotted and recalculated from previous work for comparison[21].

**Figure 4.** Plots of the 300 nm-thick Al/ZnO/ZnO$_{1-x}$/ZnO/Al based memristor: **(a)** log-log for I-V, **(b)** ln I-V $^{½}$ for SE and **(c)** ln (I/V)-V $^{½}$ for P-F. The plots of the 300 nm-thick Al/ZnO/Al based memristor: **(d)** log-log for I-V, **(e)** ln I-V $^{½}$ for SE and **(f)** 1n(I/V)-V $^{½}$ for P-F. (Solid red lines show linear fitting)

There are three different types of conduction mechanisms that generally describe the non-linear portion of the I-V curves: the Schottky emission (SE), the Poole-Frenkel (P-F) effect and the space charge limit current (SCLC), respectively [15,26–28]. The relationships of the I-V curves for the SE, PF and SCLC mechanisms are presented in **Table 3** [15,24].

**Table 3**. The expressions of SE, PF and SCLC and the associated I-V scales [21].

| Conduction Mechanism | SE | P-F | SCLC |
|---|---|---|---|
| Expressions of Mechanism | $\ln I \ \alpha \ \dfrac{e \ \sqrt{(eV)/(4\pi\varepsilon_r\varepsilon_0 \ d)}}{kT}$ | $\ln \dfrac{I}{V} \ \alpha \ \dfrac{e \ \sqrt{(eV)/(\pi\varepsilon_r\varepsilon_0 \ d)}}{kT}$ | $I \ \alpha \ \varepsilon_r\varepsilon_0\mu V^2$ |
| Associated Scales | $\ln I - \sqrt{V}$ | $\ln I/V - \sqrt{V}$ | $I - V^2$ |

V: voltage, I: current, k: Boltzmann's constant, d: active layer thickness, $T$: temperature, $\varepsilon_0$: permittivity of free space, $\varepsilon_r$: relative dielectric constant and $e$: electronic charge

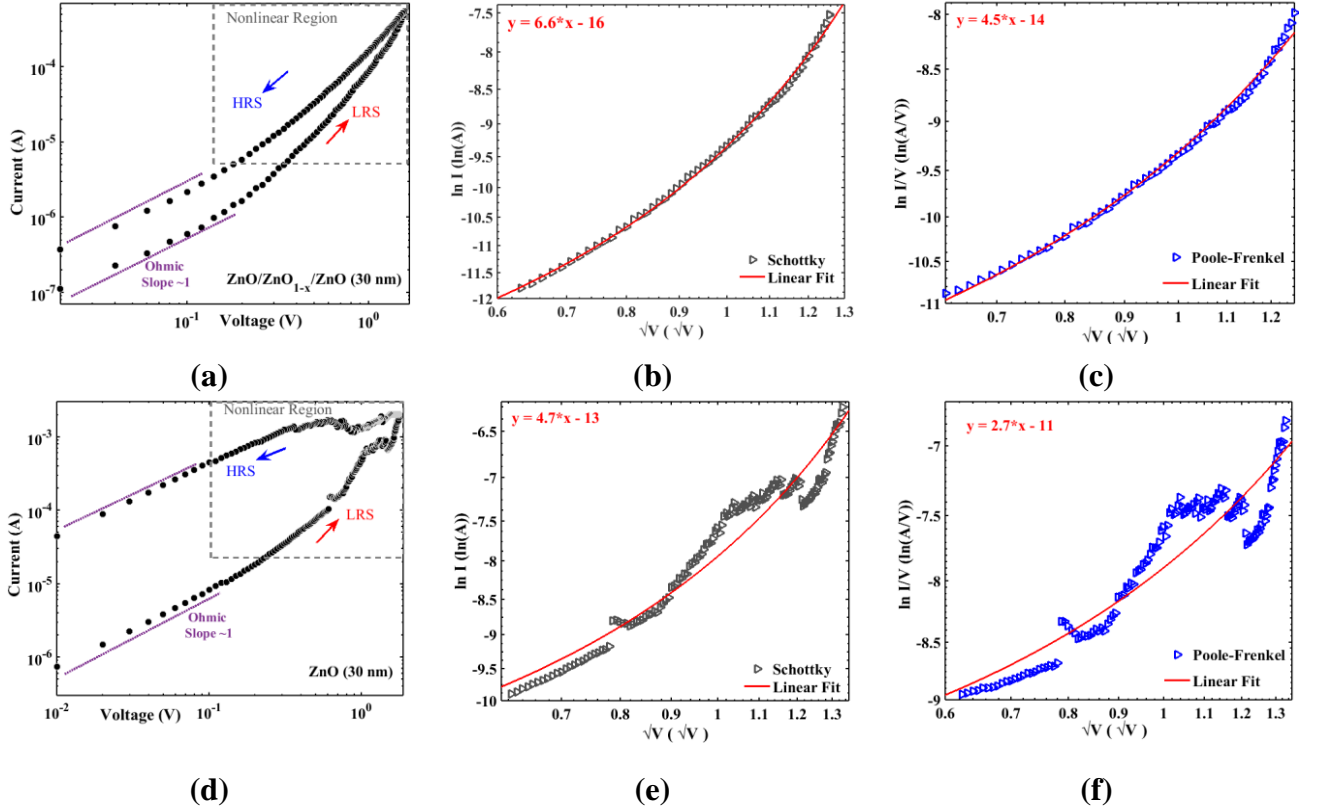

**Figure 5.** Plots of the 30 nm-thick Al/ZnO/ZnO1-x /ZnO/Al based memristor: **(a)** log-log for I-V, **(b)** ln I-V $^{½}$ for SE and **(c)** ln(I/V)-V $^{½}$ for P-F. Plots of the 30 nm-thick Al/ZnO/Al based memristor: **(d)** log-log for I-V, **(e)** ln I-V $^{½}$ for SE and **(f)** ln(I/V)-V $^{½}$ for P-F. (Solid red lines show linear fitting)

The I–V curves confined to the non-linear region were re-plotted for $\ln I\ vs.\ \sqrt{V}$ and $\ln I/V\ vs.\ \sqrt{V}$. The linear fitting lines were plotted and the linearization slope of each line was calculated (**Figure 4 (b),(c),(e),(f)** and **Figure 5 (b),(c),(e),(f)**). Since all devices showed very poor SCLC characteristics, those graphs were not included. The degree of linearity of the SE and P-F mechanisms are shown in **Table 2.** The most dominant conduction mechanism at the high electric field region of all memristor devices was the SE, followed by the P-F, in contrast to findings for ZnO-based devices in the literature [15,18,22,29]. The conduction mechanism most commonly observed in oxides is the Schottky emission [15]. Both the P-F and SE mechanisms can be caused by oxygen vacancies. The P-F mechanism may be due to the trap levels caused by oxygen vacancies inside the metal-oxide ZnO layer, while the SE mechanism may be attributed to the oxygen vacancies close to the metal/ZnO interface [24,30,31]. At the same thickness, the linearization slope of both the SE and P-F mechanisms increased as oxygen deficiencies increased.

The on/off ratio and the resistance of the HRS/LRS are some of the most important parameters when memristors are used as RS devices. **Figure 6** presents a box-plot of the HRS and LRS resistances of each device. The median-resistance values of the states, as seen in **Table 2,** are shown by the solid lines in the box. The distribution of all of the HRS and LRS resistance values was reasonably acceptable. The

30 nm-thick ZnO based device exhibited the best on/off ratio (54.8) which was more than 10 times that of the other devices. The oxygen deficient $ZnO_{1-x}$ device having the same thickness achieved an on/off ratio value of only 3.48. In addition, the 300 nm-thick devices exhibited relatively lower on/off ratios of 2.05 and 4.8 for the ZnO based and the $ZnO_{1-x}$ based devices, respectively.

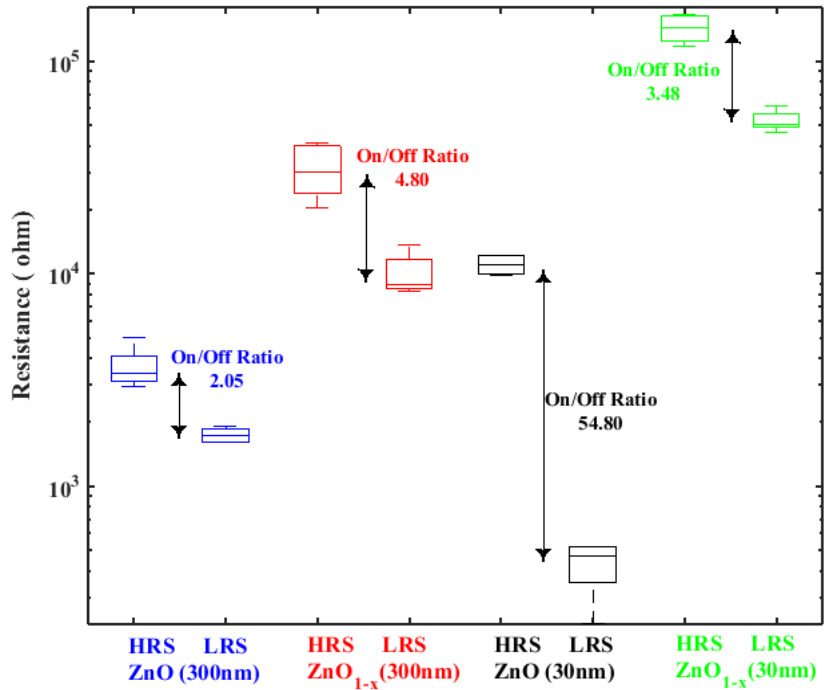

**Figure 6.** Box-plot of the HRS and LRS resistances of the ZnO and $ZnO_{1-x}$ based RS memristor devices.

Another criterion for the RS devices is the power consumption related to the compliance current. The presented devices containing intentional oxygen vacancies required nearly six times less CC, regardless of thickness, as demonstrated in **Table 2**. It was possible to obtain the lower power requirement of the RS devices by controlling the level of the oxygen vacancies.

## 4. Conclusions

In this study, 30 nm- and 300 nm-thick Al/ZnO/Al and Al/ZnO/$ZnO_{1-x}$/ZnO/Al based memristor devices were successfully produced. The pinched hysteresis I-V loops typical of memristive behavior were observed in all of the devices. The effect of cutting the $O_2$ gas flow in the middle of the process was verified experimentally via SIMS. The effect was also confirmed via analysis of the conduction mechanisms, which showed that the SE and P-F mechanisms changed in conjunction with the oxygen vacancies. This suggested that the forming process may have been related to oxygen deficiency. In addition, compliance current-dependent power requirements also decreased as oxygen deficiencies increased, regardless of the thickness of the devices. The 30 nm-thick ZnO based device displayed the best on/off ratio. It was concluded that the filamentary-switching mechanism increased as oxygen vacancies and thickness increased, while on the contrary, the homogeneous mechanism increased as oxygen vacancies increased in the thinner devices.

**Acknowledgment**

This work was partially supported by the Scientific and Technology Research Council of Turkey, under Grant No. 111T217.